*Article*

# Enhancing the Study of Quantal Exocytotic Events: Combining Diamond Multi-Electrode Arrays with Amperometric PEak Analysis (APE) an Automated Analysis Code

Giulia Tomagra [1], Alice Re [2], Veronica Varzi [2,\*], Pietro Aprà [2], Adam Britel [2], Claudio Franchino [1], Sofia Sturari [2], Nour-Hanne Amine [2], Remco H. S. Westerink [3], Valentina Carabelli [1] and Federico Picollo [2]

[1] Department of Drug and Science Technology, NIS Interdepartmental Centre, University of Torino, Corso Raffaello 30, 10125 Torino, Italy; giulia.tomagra@unito.it (G.T.); claudio.franchino@unito.it (C.F.); valentina.carabelli@unito.it (V.C.)
[2] Department of Physics, NIS Interdepartmental Centre, University of Torino and Italian Institute of Nuclear Physics, Via Giuria 1, 10125 Torino, Italy; pietro.apra@unito.it (P.A.); a.britel@unito.it (A.B.); sofia.sturari@unito.it (S.S.); nourhanne.amine@unito.it (N.-H.A.); federico.picollo@unito.it (F.P.)
[3] Neurotoxicology Research Group, Division of Toxicology, Institute for Risk Assessment Sciences (IRAS), Faculty of Veterinary Medicine, Utrecht University, P.O. Box 80.177, NL-3508 TD Utrecht, The Netherlands; r.westerink@uu.nl
\* Correspondence: veronica.varzi@unito.it

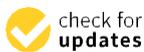

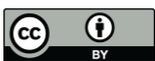

**Abstract:** MicroGraphited-Diamond-Multi Electrode Arrays (µG-D-MEAs) can be successfully used to reveal, in real time, quantal exocytotic events occurring from many individual neurosecretory cells and/or from many neurons within a network. As µG-D-MEAs arrays are patterned with up to 16 sensing microelectrodes, each of them recording large amounts of data revealing the exocytotic activity, the aim of this work was to support an adequate analysis code to speed up the signal detection. The cutting-edge technology of microGraphited-Diamond-Multi Electrode Arrays (µG-D-MEAs) has been implemented with an automated analysis code (APE, Amperometric Peak Analysis) developed using Matlab R2022a software to provide easy and accurate detection of amperometric spike parameters, including the analysis of the pre-spike foot that sometimes precedes the complete fusion pore dilatation. Data have been acquired from cultured PC12 cells, either collecting events during spontaneous exocytosis or after L-DOPA incubation. Validation of the APE code was performed by comparing the acquired spike parameters with those obtained using Quanta Analysis (Igor macro) by Mosharov et al.

## 1. Introduction

For a long time, carbon fibre electrodes (CFE) have been used as an exclusive technique for studying quantal release in neurosecretory cells [1–3]. The applicability of this technique, however, is limited by the low number of cells that can be acquired in each test; for each measurement, a new change in the fiber surface is required when moving from one cell recording to the next. Another important limitation is the need to couple the electrode with a micromanipulator to approach the cell for the measurement. Furthermore, CFEs have a short lifetime [4]. To address these limitations, over the past decade, we have developed prototypes of multi-electrode arrays known as MicroGraphited-Diamond-Multi Electrode Arrays (µG-D-MEAs) [5–7]. These prototypes showcase the capability to simultaneously capture quantal exocytotic events in real-time from both neurosecretory cells and midbrain neurons cultured directly on µG-D-MEAs. Notably, these sensors are among the few devices that enable the simultaneous analysis of amperometric measurements across multiple channels. Remarkably, these devices have shown consistent performance over more than 2 years after daily measurement sessions. The durability and reliability of the device can be attributed to the use of a diamond substrate, providing robustness, transparency, biocompatibility, chemical inertness, and resistance to biofouling.

Among the different approaches to detect exocytosis developed over the years, the most widespread is based on the amperometric detection of released oxidizable neurotransmitters. This technique takes advantage of an electrode, polarized at an appropriate bias with respect to the cellular medium, placed close (in contact) to the cell membrane to measure the current generated by the transfer of electrons following the oxidation of the molecule directly on the sensing electrode.

Amperometry provides a real-time measurement of the exocytotic events, with the limitation that it can only reveal electrochemically oxidizable molecules, such as catecholamines [8,9]. This technique has a high sensitivity, allowing the estimation of released molecules by individual vesicles since the intensity of the signal is proportional to the collected electrons. The high temporal resolution allows the quantification of the spike kinetic parameters [10]. The main steps of vesicular exocytosis after packing monoamines into vesicles begin with docking to the plasma membrane, vesicle priming and the formation of fusion pores. This is followed by fusion pore expansion and the subsequent phase of ejection of the contents towards the extracellular environment [11]. Finally, it ends with the closure of the pore [12].

In Figure 1, we show the following consecutive steps:
1. Translocation of the vesicles to the release site;
2. Dock to the plasma membrane to start the secretion molecules;
3. Membrane fusion to eject the vesicular content [12,13].

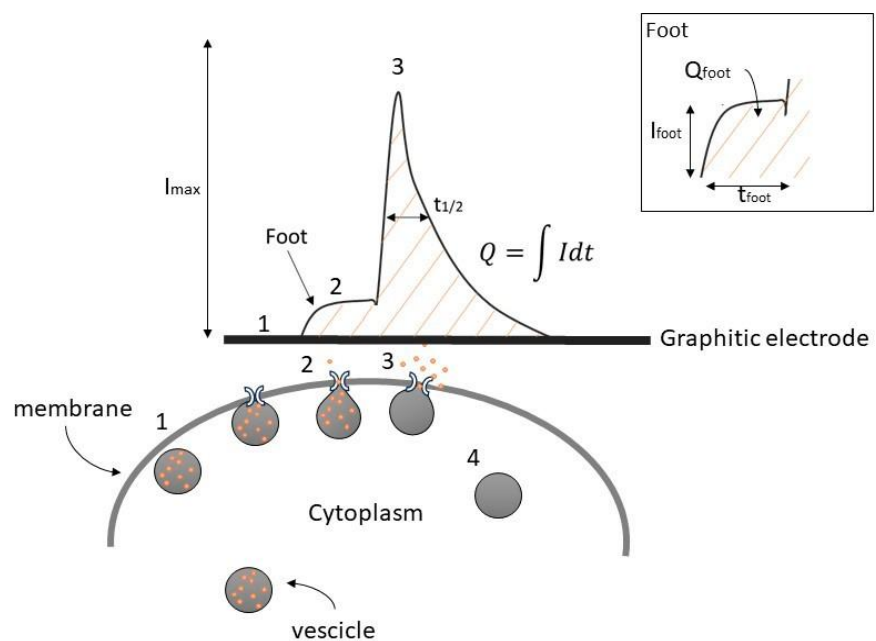

**Figure 1.** Simplified example of catecholamine secretion and its amperometric detection, showing the amperometric peak with a pre-spike foot and its descriptive parameters. Numbers indicate the progressive steps of exocytosis.

Upon the completion of exocytosis, synaptic vesicles are recovered via endocytosis and replenished with neurotransmitters (4). Alternatively, vesicles are locally recycled (kiss-and-run) or refilled with neurotransmitters without undocking (kiss-and-stay) [14–16].

In some instances, a distinctive characteristic that has been observed is a small shoulder before the amperometric spike, commonly referred to as the "pre-spike foot" (PSF) or, more simply, foot. The exact nature of this phenomenon during the exocytotic process remains a subject of ongoing debate and study [16–18]. Historically, the onset of exocytotic events is described as a gradual release of a small number of catecholamine molecules from a narrow "fusion pore" having a nanometric dimension before the pore itself undergoes dilation for complete exocytosis [19,20]. The conditions of foot formation are associated with the instability of fusion of the granules with the cell membrane, which in the specific condition is not permanent and therefore the fusion pore opens and closes reversibly, so-called flickering fusion [21].

According to Amatore et al. [22], the foot represents the release of a small fraction of the molecules stored in vesicles. In chromaffin cells, it has been observed that the proportion of spikes with a foot remains consistently around 30% in each experiment [20,22,23]. Therefore, examining the relevant pre-spike parameters is essential for comprehending the dynamics of the fusion pore that remain the subject of continued interest and ongoing debate.

The key parameters for a comprehensive description of exocytotic spikes (see Figure 1) include the maximum oxidation current ($I_{max}$), the half-time width of the spike evaluated at 50% of $I_{max}$ ($t_{1/2}$), the total charge detected (Q), corresponding to the area under the peak which is related to the number of oxidized molecules (*N*) by a single vesicle following the equation:

$$N = Q/(n \cdot e)$$

where e is the elementary charge ($1.6 \cdot 10^{-19}$ C), and *n* is the number of electrons transferred during the oxidation of a single molecule to the sensitive electrode [24].

Similarly, the foot can be quantified by the following parameters: $I_{foot}$ (foot current, defined as the mean current measured for the duration of the foot or the plateau current when a steady state can be observed), $t_{foot}$ (foot duration), and $Q_{foot}$ (area under the foot) [25].

Given the importance of spike parameters quantification, having implemented a sensing multi-array that allows the simultaneous spike detection from 16 electrodes, it was essential to develop a code that could support a reliable, fast, and simultaneous analysis of signals deriving from all electrodes.

In this study, we present an improved approach for the effective investigation of the quantal exocytotic events using the combination of diamond-based multi-electrode arrays with graphite microelectrodes (µG-D-MEAs) and APE (Amperometric Peak Analysis) code. This newly developed analysis software, written in MATLAB (R2022a), enables a swift and automated analysis of amperometric spike parameters from multiple simultaneous amperometric recordings (chronoamperograms).

Presently, due to the absence of commercial multi-electrode arrays supporting amperometric measurements, dedicated software for concurrent analysis across different channels on the same device is lacking. Therefore, researchers must conduct the analysis channel by channel, consuming a substantial amount of time due to the significant effort required. This necessity prompted the creation of the APE software (http://www.solid.unito.it/Software/Software_index.html, accessed on 10 November 2023), facilitating the completion of analyses for several electrodes within a few minutes.

Furthermore, the APE software distinguishes itself from existing solutions by eliminating the need for experimenters to manually input threshold or background parameters. Instead, it autonomously generates and provides the necessary parameters for analysis. This feature represents a significant advantage, promoting a standardized analysis that is independent of individual experimenter inputs.

Amperometric measurements were performed using PC12 cells, collecting signals from several individual cells during each experiment (up to 16, corresponding to the number of independent electrodes). Trials include monitoring the spontaneous exocytotic activity under control conditions and after L-3,4-dihydroxyphenylalanine (L-DOPA) [26] administration. Acquired data were analyzed using the APE code and compared to Quanta Analysis (Igor macro) by Mosharov et al. [10] to validate the developed software.

## 2. Materials and Methods

### 2.1. MicroGraphited-Diamond-Multi Electrode Arrays (µG-D-MEAs)

In recent years, we have developed diamond-based biosensors with multiple sub-surface graphitic channels used as sensitive electrodes for electrochemical detection schemes, so-called MicroGraphited-Diamond-Multi Electrode Arrays (µG-D-MEAs).

These devices take advantage of the specific properties of the two carbon allotropes: diamond offers a perfect matrix for the development of cellular sensing due to its good biocompatibility [27,28], as well as its high optical transparency, essential for both standard transmission microscopy (observation of cells) and fluorescence microscopy. Meanwhile, the electrical properties of graphite ensure a lower resistivity [29], wide electrochemical windows and an optimal compromise for the surface capacity, allowing the employment of this material as an electrode for amperometric or potentiometric recording of biosignals.

µG-D-MEAs are fabricated via the ion beam lithography technique [30]. This technique exploits the metastable nature of the diamond, which promotes its conversion into amorphous carbon, induced by the accumulation of the vacancy created by the interaction of carbon atoms with the ion beam. In fact, after overcoming a critical threshold called graphitization threshold (in the range comprised between $1 \times 10^{22}$ and $9 \times 10^{22}$ cm$^{-3}$ [31]), the diamond lattice loses its original structure and becomes a random network of carbon atoms with $sp^2$ and $sp^3$ hybridization. After thermal annealing at high temperature (>900 °C), the amorphous structure is transformed into graphite, which is the thermodynamically stable allotrope, while point defects (i.e., Frenkel defects) are recovered, restoring the original diamond structure.

The implementation of this process with MeV ions makes it possible to take advantage of the non-uniform distribution of the induced vacancies, which are mainly created in correspondence with the Bragg peak located in the bulk of the implanted target, and thus to promote graphitization up to several tens of microns below the diamond surface.

The implantation is carried out using high-resolution masks that define the 3D geometry of the graphitic structures, which are completely embedded in the diamond matrix and only emerge at the surface at their ends, allowing them to be connected to the cells in the centre of the biosensor and to the front-end electronics at the periphery.

Ion irradiation was performed at room temperature at the INFN National Laboratories of Legnaro [32] with a broad beam of light ions (i.e., He) with energy comprised between 0.5 MeV and 2 MeV.

The biosensors fabricated have 16 independent electrodes for the real-time and simultaneous amperometric detection of the exocytotic events occurring from many cells and an Ag/AgCl quasi-reference electrode which is placed in the extracellular medium. The signals are digitally converted by a USB-6216 ADC module (National Instruments, Austin, TX, USA) setting a sampling rate at 25 kHz. The bias applied to the channels is +800 mV, a value that guarantees the oxidation of dopamine with an optimal signal-to-noise ratio [33–35].

### 2.2. PC12 Culture Cells

A rat pheochromocytoma (PC12) cells subclone [36,37] was kindly provided by R. Westerink [38] and maintained in flasks in an incubator set at 37 °C with an atmosphere of 5% $CO_2$ and 95% humidity [39]. This cell line shares both the secretory mechanism and the neurotransmitters released with dopaminergic neurons, making them a versatile and robust biological model for exocytosis experiments [35,36].

Interfacing PC12 cells with µG-D-MEAs requires the following steps:

First, µG-D-MEAs have to be coated with collagen type I (Sigma-Aldrich, St. Louis, MO, USA) and then placed in the incubator for 4–5 h; afterwards, the µG-D-MEAs are gently rinsed with phosphate-buffered saline (PBS) to remove the excess of collagen.

Next, PC12 cells are detached from the flask in which they are grown using trypsin EDTA (0.25% Sigma-Aldrich) for 6–8 min in the incubator. After this step, PC12 cells are centrifuged (1200× *g* for 4 min) and then resuspended in a culture medium for seeding on the sensor surface at a density of ✕ $10^6$ cells $cm^2$. Cells are maintained in RPMI-1640 (Invitrogen, Waltham, MA, USA) medium (containing 10% horse serum (Invitrogen), 5% fetal bovine serum (Invitrogen) and 2% antibiotic/antimitotic (pen/strep Invitrogen), at a temperature of 37 °C in a 5% $CO_2$ atmosphere. The experiments are performed at room temperature within 2–4 days after culturing without changing the culture medium.

For a subset of experiments, PC12 was incubated with Levodopa (L-DOPA, 20 µM, Sigma Aldrich) for 30 min before the experiment and maintained at 37 °C in a 5% $CO_2$ humified atmosphere.

*2.3. APE Code Structure*

Multi-electrode arrays developed for amperometric recordings allow the simultaneous recording of multiple chronoamperograms, each consisting of tens/hundreds of exocytotic events. These large amounts of data require rigorous analysis to extract the relevant spike parameters (maximum current amplitude, charge, number of molecules, kinetic parameters) and therefore dedicated software to perform automated and flexible procedures.

APE code (Amperometric Peak Analysis) is a freeware analysis code written in MATLAB which allows a fast and automated analysis of amperometric peak parameters from different chronoamperograms. This code has been designed to be user-friendly, with a clean interface for data entry and storage of output reports in ASCII, allowing high compatibility with other analysis software. The methods for extracting spike parameters are compatible with strategies proposed in the literature and have already been implemented in other software [10].

Chronoamperograms are characterized by a noise bandwidth of 5–8 pA, while amperometric spike peaks appear exclusively as positive (upwards-directed) signals. In order to correctly identify the signals and eliminate artifacts, a detailed iterative procedure is carried out by the developed software.

- Data import

Data are loaded in .txt format, and the data matrix is created. Once the data have been loaded, a window appears in which it is possible to insert the value of the sampling frequency. Thus, the time variable array is generated, as it may not always be directly available in the sensor output. It is determined by dividing the number of data points by the sampling frequency. Once the sampling frequency is inserted, a second window allows the setting of the maximum spike duration; this value is used to discharge multiple convoluted signals due to the detection of simultaneous exocytotic events. This parameter is not uniquely determined; rather, it varies depending on the specific cellular culture being analyzed. Future versions of the code will identify and separate convolved signals without the need for external parameters to be inserted by the user.

- Front-end electronic gain

The user is asked to insert the multiplication factor of the own amplification system (if it is foreseen; if not, this parameter must be set equal to 1).

- Channel selection

The user must select the channel(s) to be analyzed by selecting the corresponding windows, as shown in Figure 2. All other channels are not stored to save memory space and speed up the subsequent analysis steps. This step could be useful to exclude from the analysis silent channels or damaged channels (i.e., high noise band) that are not acquiring any significant signals. For example, in Figure 2, channels #2, #10 and #11 display noisy tracks (note the difference in scaling), which were not included in the analysis. The data from all selected channels are analyzed simultaneously, highlighting a key advantage of the APE software. Unlike other methods, APE allows for the concurrent analysis of channels rather than individual channel analyses.

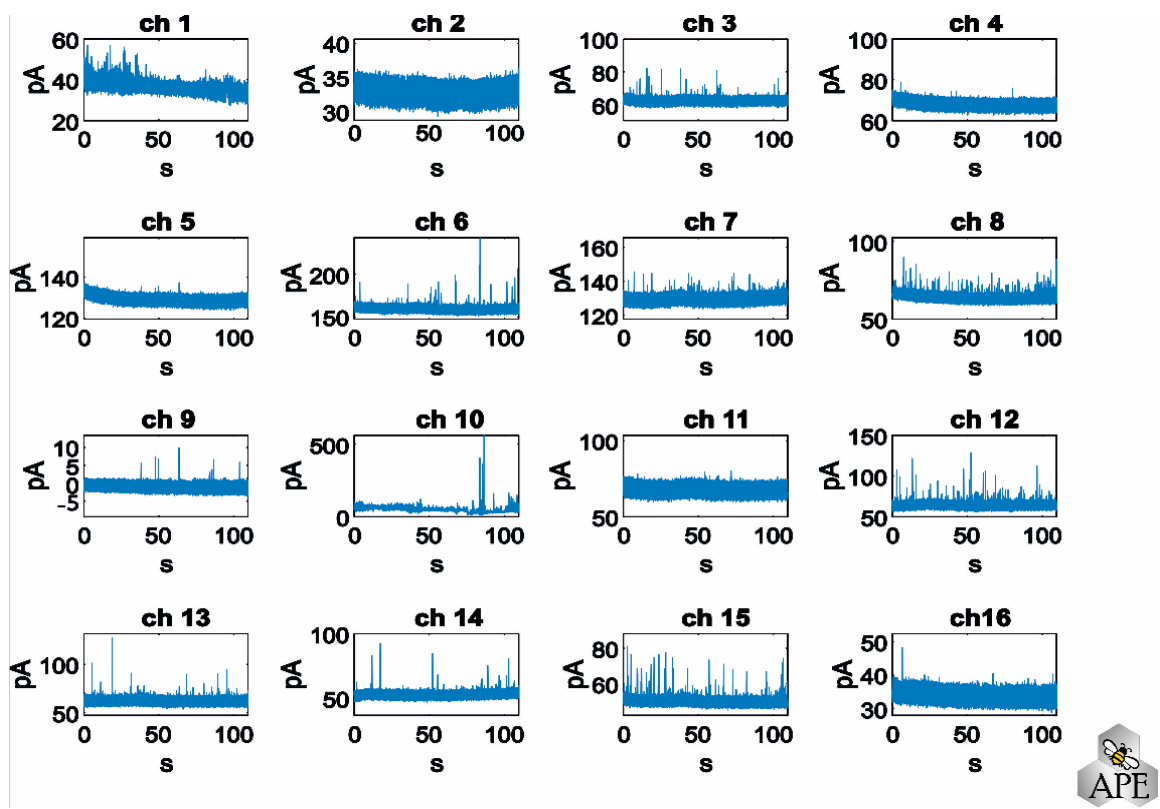

**Figure 2.** Representative chronoamperograms recorded simultaneously from 16 electrodes.

- Signal filtering

The data are filtered to optimize the peak identification performed in the next step. The chosen filter is the PT1 filter [40]. This filter is a first-order low-pass filter, representing a continuous-time linear system that attenuates or amplifies an input signal depending on its frequency. The PT1 filter exhibits a time-decreasing exponential impulse response, signifying that the output signal gradually converges towards the input signal over time. Implementation of this filter sets the noise baseline to a zero value.

- Automatic threshold determination and peak sorting

The identification of the signal with respect to the baseline is determined by a thresholding process.

Assuming the background noise follows a white noise pattern with a Gaussian distribution, the threshold is established as a multiple of the standard deviation ($\sigma$). While users have the flexibility to adjust the multiplication factor applied to the standard deviation, optimal spike sorting typically employs a setting of 3 $\sigma$, encompassing 99% of the noise. In Figure 3, three scenarios are depicted where the multiplication factor is set to low (1 $\sigma$), optimal (3 $\sigma$), and high values (10 $\sigma$), representing, respectively, the misidentification of noise as a signal, the identification of all amperometric peaks, and the loss of numerous spikes.

- Peak analysis and parameters extraction

In order to quantify the amperometric spike parameters, a baseline is established for each spike by fitting a linear model to the data that spans from the spike's beginning to its end. The boundaries of these spikes are determined by comparing the signal's amplitude to the statistical mode of the background signal in close proximity to the amperometric spike.

This comparative analysis is conducted both forwards and backwards, centring around the location of the previously identified peak's apex. If the signal amplitude at a specific point surpasses the statistical mode of the background, it is recognized as part of the spike.

Conversely, the first point that fails to meet this criterion is designated as the starting or ending point of the spike.

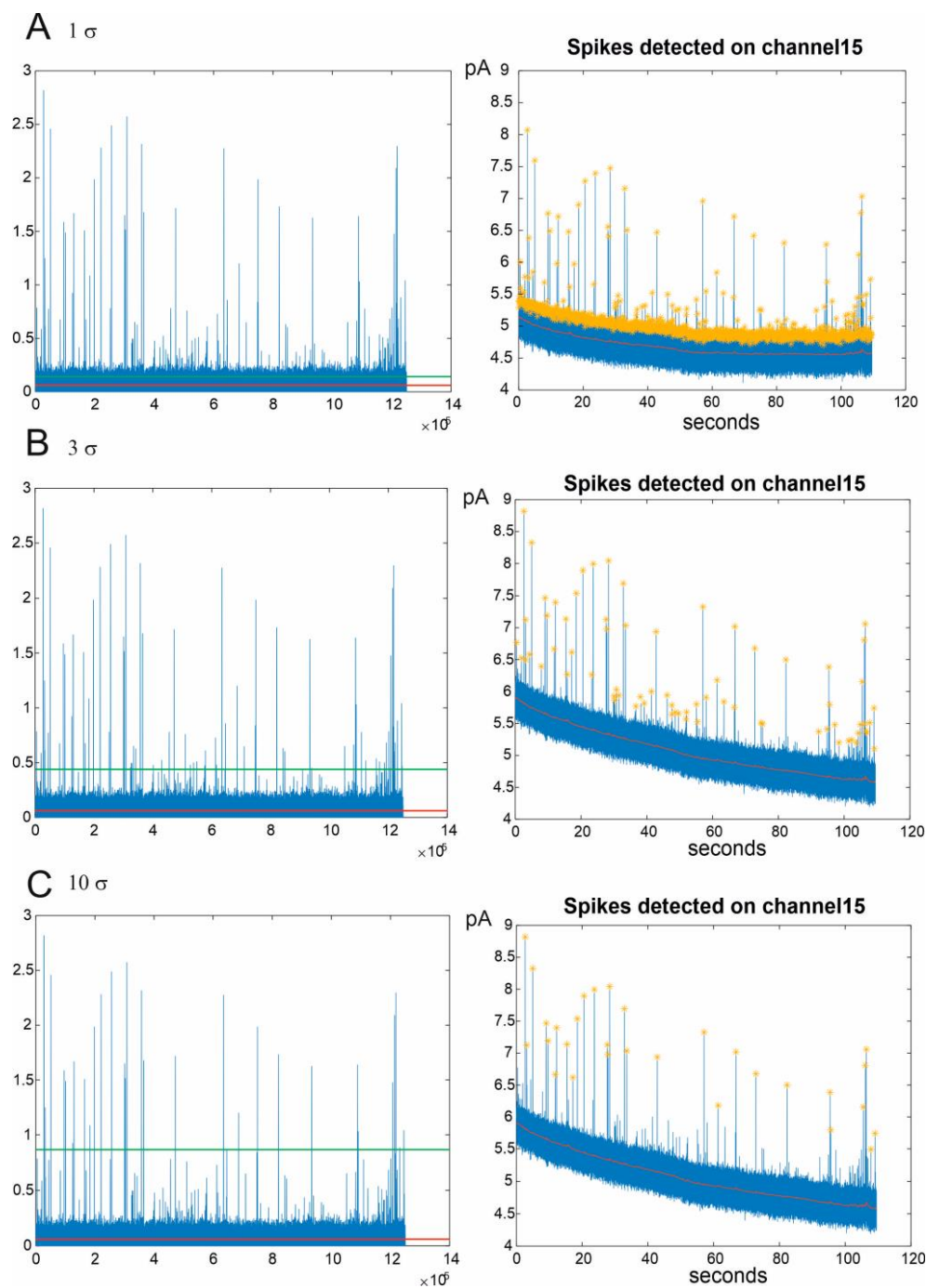

**Figure 3.** Amperometric signal identification in three different configurations of spike sorting: 1 σ (**A**), 3 σ (**B**), and 10 σ (**C**). (**Left**): signals filtered through the PT1 filter. The red line indicates the subtraction of the filtered background. The green line represents the threshold above which the amperometric spikes are identified. (**Right**): chronoamperograms of the spikes detected by a representative channel. The yellow stars indicate the detected spikes, while the red line is the background.

This systematic approach ensures the precise extraction of amperometric spike parameters and facilitates a comprehensive understanding of the signal characteristics. The procedure for the evaluation of the spike parameters is given below:

- The **maximum current** of the spike $I_{max}$ (pA) is determined by the difference between the maximum amplitude of the spike and the corresponding point of the baseline.
- The **duration** (ms) of the spike is evaluated as the time difference corresponding to the beginning and end of the spike peak.
- The **total charge** Q (pC) is evaluated as the area under the spike, subtracting the area under the baseline. From the total charge value, the number of secreted molecules is evaluated considering the number of electrons that the molecules release to the electrode during the oxidation reaction (the user can choose the number of oxidation electrons related to the catecholamine of interest, 2 for dopamine [41]).
- $t_{1/2}$ (ms) corresponds to the first point whose corresponding current value is equal to or less than half of $I_{max}$ (the search starts from the peak maximum and goes forwards and backwards until such a point is found).
- The **slope** of the rising phase of the peak is obtained via a linear fit. The values used in the program correspond to 25% of $I_{max}$ and 75% of $I_{max}$ on the ascending part of the spike. The difference between the occurrence time of the values used is then stored as the rise time of the peak.
- The **falling phase** of the spike is evaluated by fitting the dataset with double or single exponential functions. Each spike is translated so that its starting time is zero, and the baseline (as always obtained by a linear fit) is subtracted from the data. After this normalization, the data between 75% of $I_{max}$ and the end of the spike are fitted. The algorithm compares the SSE (summary squares error) of the fits (single or double exponential) and selects the one that best represents the data of each peak.

- Foot analysis and parameters extraction

There is also a dedicated algorithm to detect the pre-spike foot in the amperometric peaks, which is characterized by many of the features used to describe the spike listed in the previous steps of the analysis. The routine for the foot identification compares the current values before the end of the foot (see below) with the average of the signal noise. If a significant number of points ($n_{point}$ = 40) are found whose height is greater than the noise average, the last point greater than the average is stored as the start of the foot. The end time of the foot is evaluated as the intersection of the baseline and the rise phase of the spike. The foot charge is evaluated as the area under the points found.

## 3. Results

Experiments were performed using µG-D-MEAs to collect PC12 cells' spontaneous exocytotic activity and also to investigate the effect of L-DOPA.

Cells were cultured on µG-D-MEAs, as described in the Section 2, removed from the $CO_2$ incubator, and placed in the front-end electronics, allowing the signals to be collected. The typical duration of the amperometric recordings was 120 s. Figure 4 shows the µG-D-MEA and a representative chronoamperogram highlighting two examples of different signals: an amperometric spike in which the foot is clearly detectable (left, C) and a second one without foot (right, D).

The recorded spikes were analyzed using the APE code, and obtained parameters were compared with those derived from the widely used Quanta analysis macro (based on IGOR Pro 8 software) developed by Mosharov et al. [10] since it represents a benchmark to validate the effectiveness of our analysis software.

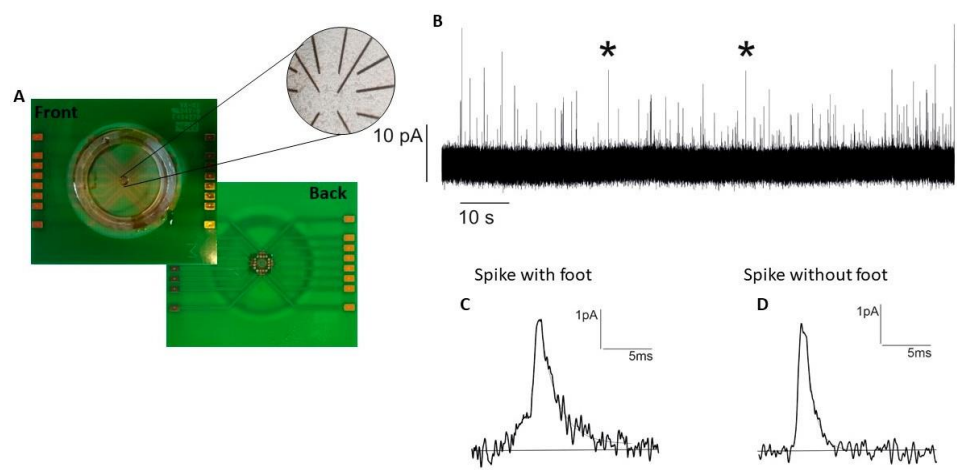

**Figure 4.** (**A**) µG-D-MEAs and PC12 cells image with the graphitic tracks shown at an enlarged scale. (**B**) Representative amperometric trace obtained from one of the 16 electrodes. (**C**,**D**) Examples of amperometric peaks with (**C**) and without (**D**) foot. These examples are denoted with an asterisk in (**B**).

*3.1. APE Code Validation*

The results obtained from the APE code analysis were compared with those obtained from the manual analysis using a Quanta Analysis macro in order to validate the correct functioning of APE code.

The time required by the experimenter to perform the analysis with the APE code, using a mid-range laptop (Lenovo AMD Ryzen 5000, 16 Gb RAM, Morrisville, NC, USA), was ~10 min for each biosensor; the analysis is conducted simultaneously for all electrodes (for 16 electrodes. Using IGOR software, the manual analysis required is approximately between 30 min and 1 h for each electrode (total recording trace: 120 s) for an expert user; the analysis is conducted one electrode at a time. The creation of the data matrix represents the longer step in time of the routine while the extraction of the spikes parameters requires just a few seconds.

We compared the parameters of single amperometric spikes using three different software: IGOR Pro 8 (Quanta analysis macro), APE code and Origin Pro 2022 (advanced data analysis software). The obtained parameters are compatible regardless of the software used, as shown in Figure 5 for a single representative spike.

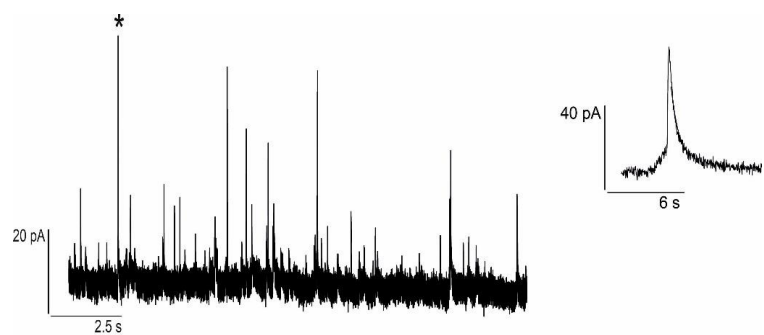

|  | $I_{max}$ [pA] | $t_{1/2}$ [ms] | Q [pC] |
|---|---|---|---|
| IGOR | 58.9 | 4.6 | 0.52 |
| APE | 59.1 | 4.5 | 0.54 |
| Origin pro | 58.0 | 4.4 | 0.56 |

**Figure 5.** Example of a single amperometric spike (denoted with an asterisk in the trace on the (**left**) and depicted in detail on the (**right**)) analysis involving three different software: IGOR Pro 8, APE, Origin Pro 2022.

In Figure 6, we report the comparison of Q, $t_{1/2}$, $I_{max}$, and spike frequency analyzed by means of the APE code and with the software Quanta analysis written in Igor. Spikes were measured in the control condition (culture medium, see Section 2) ($n_{spikes}$ = 427, $n_{channels}$ = 8) and after L-DOPA incubation ($n_{spikes}$ = 536, $n_{channels}$ = 6). As shown in Figure 6, there are no significant differences for any of the reported parameters (Q, $t_{1/2}$, $I_{max}$, and frequency) between the two programs, IGOR and APE code. Furthermore, the APE code works correctly to identify the altered spike parameter after administering a drug (L-DOPA). Also, in the presence of L-DOPA, the data analyzed with the APE code are consistent with those analyzed with IGOR software.

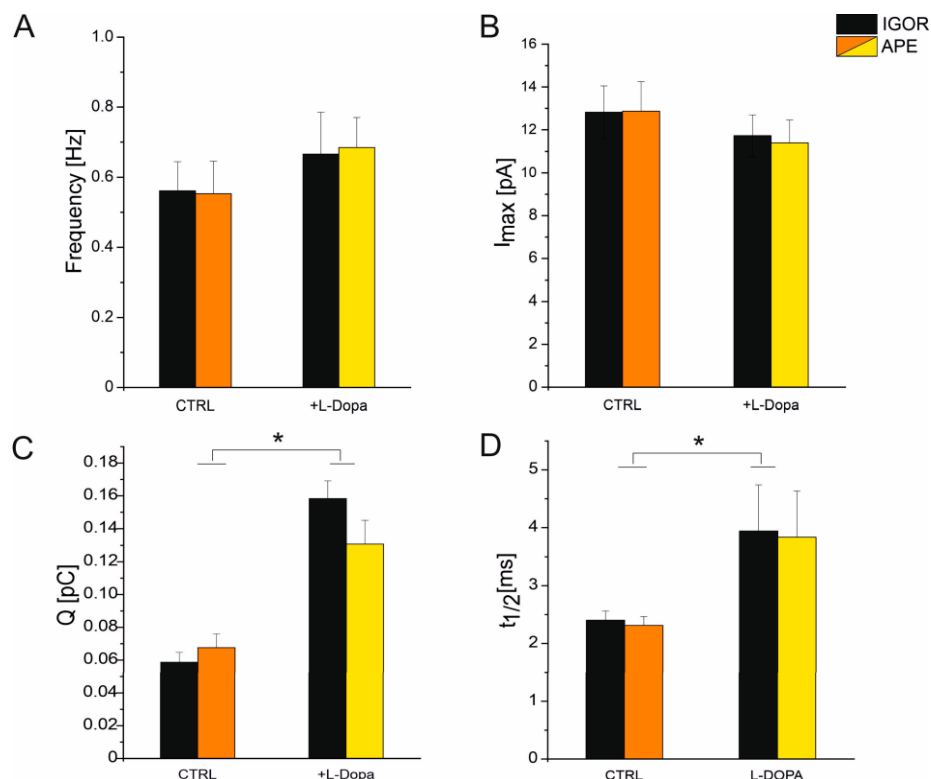

**Figure 6.** Comparison between the analysis performed using the automatic APE software and the analysis performed manually using IGOR (Quanta Analysis). Spike parameters are shown under control conditions and after incubation with L-DOPA. Black bars refer to data obtained using Igor. Orange/yellow bars refer to data obtained using APE (see legend). No significant differences ($p > 0.05$) are observed for the mean values of the parameters evaluated using the two software: (**A**) frequency, (**B**) maximum current amplitude $I_{max}$, (**C**) charge (Q), and (**D**) $t_{1/2}$. After L-DOPA incubation, Q and $t_{1/2}$ are significantly different from control ($p < 0.05$) as indicated by * [35].

These results clearly demonstrate the efficiency of the APE software in analyzing amperometric parameters in a short time in the control condition and after the drug test.

### 3.2. Case Study with µG-D-MEAs + APE Code: Foot Analysis

After confirming the capability of the APE software to properly identify and characterize the amperometric spikes, we focused on the analysis of the foot. The presence of such a "foot" preceding the full fusion event is usually attributed to the oxidation of a small amount of material released via a nanometric fusion pore between the cell membrane and the vesicle, which interact to fuse together [20].

First of all, we evaluated the number of peaks in which the foot was present. Under control conditions, we could reveal the foot in 25 ± 1% of spikes, in good agreement with Amatore et al. [20], who estimated that 30% of the amperometric peaks displayed a significant foot current.

The foot signals were also examined after incubation of PC12 cells with L-DOPA (see Section 2). This substance is the precursor of dopamine, which stimulates the cellular production of dopamine and consequently increases the content of this neurotransmitter in the secretory vesicle without affecting the frequency of exocytotic events [41–43].

After incubation with L-DOPA, as observed for chromaffin cells [22], PC12 cells also showed an increase in the frequency of amperometric spikes exhibiting a foot signal, which corresponds to 31 ± 2% of the spikes (versus 25% observed in control conditions). Among the potential hypotheses for the increment of the frequency of the observed foot, the work conducted by Amatore and colleagues [20] suggested that the identification of an amperometric foot is not solely dependent on the stability of the fusion pore and external physicochemical factors. Notably, the proportion of spikes exhibiting a foot remains consistent even under varied constraints such as curvature, tension, and viscosity applied to chromaffin cell membranes. In contrast, alterations in granule composition, such as the influence of L-DOPA or reserpine, result in changes in the frequency of detecting an amperometric foot. We can strengthen this hypothesis by confirming that it is also independent of the biological system used since it is observed not only in primary chromaffin cells but also in PC12 cells.

This implies that the presence of a pre-spike foot may be associated with the structure and composition of secretory vesicles, specifically the existence of an intravesicular medium with a diffusion coefficient for catecholamines larger than that in the densely packed core matrix.

Focusing on foot parameter analysis ($I_{foot}$, $Q_{foot}$ and $t_{foot}$), after L-DOPA incubation, a significant increase in foot-charge and foot-time was revealed, respectively from 0.014 ± 0.001 pC (ctrl $n$ = 76) to 0.026 ± 0.003 pC (L-DOPA $n$ = 100, $p$ < 0.05) and from 3.3 ± 0.2 ms (ctrl, $n$ = 76) to 5.3 ± 0.4 ms (L-DOPA $n$ = 100, $p$ < 0.05). Instead, $I_{foot}$ was not significantly altered, being 4.37 ± 0.16 pA in the control condition ($n$ = 76) and 4.48 ± 0.23 pA after L-DOPA incubation ($n$ = 100) ($p$ > 0.05) (Figure 7). These results are in good agreement with those observed by Amatore et al. [20].

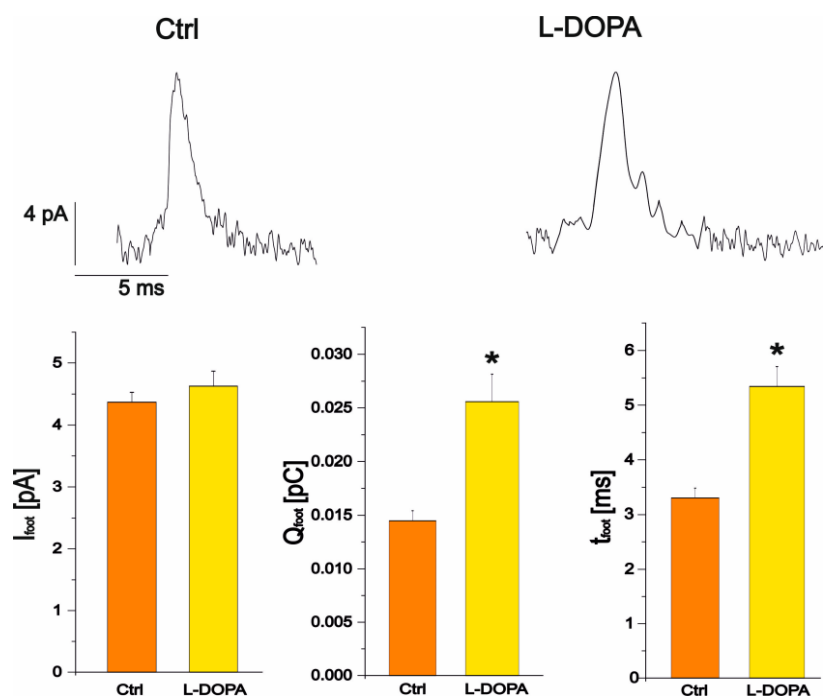

**Figure 7.** Representative spikes in the control condition (**left**) and after L-DOPA administration (**right**). The bar plots represent the maximum current of the foot in control and, after L-DOPA incubation ($p$ > 0.05), the charge of the foot and the duration of the foot. Charge and duration of the foot are significantly increased by L-DOPA compared to control ($p$ < 0.05) as indicated by *.

## 4. Conclusions

The combination of a multi-electrode array dedicated to amperometric measurements with fast and effective analysis software will present a significant improvement in the study of quantal secretory events.

The developed µG-D-MEAs are electrochemical sensors with a high signal-to-noise ratio [35], and a good temporal resolution (<ms) that allow the detection of fast exocytotic events as well as the identification of pre-spike foot. On the other hand, the collected data need appropriate tools to be analyzed: for example, when acquiring data from many channels, it is mandatory to speed up the process and reduce the bias due to the experimenter, improving the automated routine for signal sorting and parameter evaluation, as performed by the APE code.

The PC12 cell line was chosen to carry out this study by detecting and measuring both the amperometric spikes under spontaneous exocytotic activity and also the related foot events. In line with what has been observed by different research groups [13,20,39] studying foot in chromaffin cells, we have shown that foot is present in $25 \pm 1\%$ of cells also in PC12 cells and that the percentage of foot significantly increases after L-DOPA administration ($31 \pm 2\%$, $p < 0.05$).

The increased frequency of foot signals after L-DOPA administration is likely to be due to the increased vesicle volume following increased catecholamine content [13,44]. This effect contributes to the formation of a "halo" consisting of the space between the dense core and the vesicle membrane.

We observed that the maximum foot current amplitude, corresponding to the flux of released neurotransmitter, remained constant after L-DOPA incubation (from $4.37 \pm 0.16$ pA in control conditions to $4.48 \pm 0.23$, $p > 0.05$). Instead, there is a significant increase in the duration of the foot signal ($t_{foot}$), which represents the lifetime of the foot and indicates better stability of the fusion pore, while $Q_{foot}$, which correlates with the number of molecules released via the pore during the lifetime of the foot (Figure 7), also increases. These results are consistent with the phenomenological pathway based on an increase in the volume of the halo after incubation with L-DOPA, i.e., an increase in the catecholamine content of the halo, as observed experimentally with the increase in the foot charge [25].

The increase in $t_{foot}$ and $Q_{foot}$ after L-DOPA incubation shows a good agreement with the variation in amperometric peak parameters induced by L-DOPA administration to PC12 cells. Both $t_{1/2}$ and Q mean values significantly increased after incubation with L-DOPA compared to the control condition. These observations support the hypothesis that L-DOPA favours multi-event fusion and/or increases vesicle diameter and volume. The observation that $t_{foot}$ increases after L-DOPA incubation reinforces this theory, as more fusion pore stability is required for multiple events and/or larger vesicles.

In conclusion, to demonstrate the reliability of our software, we compared the parameters extracted with the APE code with those obtained using the Macro Quanta analysis developed by Mosharov and colleagues [10], which is widely used in the neuroscience community. Our data demonstrate that µG-D-MEAs provide an excellent multi-electrode tool that allows for the discrimination of the presence of the foot in an amperometric signal and that, combined with the APE code, guarantees a standardized and rapid analysis of the amperometric peaks and related foot parameters.

**Author Contributions:** Conceptualization, G.T. and F.P.; methodology, C.F.; software, G.T., A.R. and V.V.; formal analysis, A.R. and S.S.; investigation, G.T., A.B. and N.-H.A.; data curation, G.T.; writing—original draft preparation, G.T. and A.R.; writing—review and editing, G.T., V.V., P.A., A.B., N.-H.A., R.H.S.W., V.C. and F.P.; project administration, F.P.; funding acquisition, V.C. and F.P. All authors have read and agreed to the published version of the manuscript.

**Funding:** The "Trapezio" project was funded by the Compagnia di San Paolo (2022–2024, n. 2021.2236), the "QuantDia" project funded by the Italian Ministry for Education, University and Research (MIUR) within the "FISR 2019" program, and the "LasIonDef" project was funded by the European Research Council under the "Marie Skłodowska-Curie Innovative Training Networks" program (GA n. 956387).

**Institutional Review Board Statement:** Not applicable.

**Informed Consent Statement:** Not applicable.

**Data Availability Statement:** Data are contained within the article.

**Conflicts of Interest:** The authors declare no conflict of interest.